\documentclass[aps, prc, reprint, amsmath, amssymb, superscriptaddress, floatfix, nofootinbib]{revtex4-1}

\usepackage[pdftex]{graphicx}
\DeclareGraphicsExtensions{.jpg,.png,.pdf}

\usepackage[utf8]{inputenc}
\usepackage{amsmath}
\usepackage{amssymb}
\usepackage{amsfonts}
\usepackage{tabularx}
\usepackage{booktabs}
\usepackage{graphicx}
\usepackage{color}
\usepackage{multirow}
\usepackage[perpage,symbol]{footmisc}
\usepackage{lineno}
\usepackage{verbatim}
\usepackage{dcolumn}
\usepackage{bm}
\usepackage[inline]{enumitem}
\graphicspath{{Fig/}}
\definecolor{theblue}{RGB}{0,50,230}
\usepackage{appendix}
\usepackage{setspace}
\usepackage{hyperref}
\hypersetup{
  colorlinks=true,
  linkcolor=theblue,
  citecolor=theblue,
  urlcolor=theblue
} 

\newcommand{\pt}{\ensuremath{p}_{\perp}}

\newcommand{\snn}{\sqrt{s_{\rm NN}}}

\begin{document}


\title{Lepton pair photoproduction in hadronic heavy-ion collisions with nuclear overlap}

\author{Kewei~Yu}
\affiliation{%
College of Science, China Three Gorges University, Yichang 443002, China\\
}%
\affiliation{%
Center for Astronomy and Space Sciences, China Three Gorges University, Yichang 443002, China\\
}%

\author{Jiazhen~Peng}
\affiliation{%
College of Science, China Three Gorges University, Yichang 443002, China\\
}%
\affiliation{%
Center for Astronomy and Space Sciences, China Three Gorges University, Yichang 443002, China\\
}%

\author{Shuang~Li}
\email{lish@ctgu.edu.cn}
\affiliation{%
College of Science, China Three Gorges University, Yichang 443002, China\\
}%
\affiliation{%
Center for Astronomy and Space Sciences, China Three Gorges University, Yichang 443002, China\\
}%

\author{Kejun~Wu}
\email{wukj@ctgu.edu.cn}
\affiliation{%
College of Science, China Three Gorges University, Yichang 443002, China\\
}%
\affiliation{%
Center for Astronomy and Space Sciences, China Three Gorges University, Yichang 443002, China\\
}%

\author{Wei~Xie}
\affiliation{%
College of Science, China Three Gorges University, Yichang 443002, China\\
}%
\affiliation{%
Center for Astronomy and Space Sciences, China Three Gorges University, Yichang 443002, China\\
}%

\author{Fei~Sun}
\affiliation{%
College of Science, China Three Gorges University, Yichang 443002, China\\
}%
\affiliation{%
Center for Astronomy and Space Sciences, China Three Gorges University, Yichang 443002, China\\
}%

\date{\today}%

\begin{abstract}
Lepton pair photoproduction provides a unique probe
of the strong electromagnetic field produced in ultra-relativistic heavy-ion collisions.
To map out the broadening behavior of the lepton pair transverse momentum with respect to
the back-to-back correlation structure,
we present a theoretical model based on the equivalent photon approximation,
and then we update it to make direct comparisons with the recent experimental measurements.
We find that the model calculations can describe well,
not only the average transverse momentum squared of $e^{+}e^{-}$ pairs in Au--Au collisions at $\snn=200$~GeV,
but also the acoplanarity of $\mu^{+}\mu^{-}$ pairs in Pb--Pb collisions at $\snn=5.02$~TeV.
Furthermore, the model predictions are also able to reproduce
the measured lepton pair mass and transverse momentum squared distributions.
\end{abstract}

\maketitle


\section{INTRODUCTION}\label{sec:Intro}
In ultra-relativistic collisions of heavy nuclei such as Au or Pb,
extremely strong electromagnetic fields are generated when the two nuclei
pass through each other~\cite{Skokov:2009qp, PhysRevC.85.044907, Pu:2016ayh}.
The resulting strong electromagnetic fields provide a unique opportunity
to investigate some quantum electrodynamics (QED) effects,
such as light-by-light scattering~\cite{dEnterria:2013zqi},
which was confirmed experimentally by ATLAS~\cite{ATLASnaturePhy2017} and CMS Collaborations~\cite{CMS:2018erd},
and the matter productions (e.g. dileptons and $J/\Psi$) directly from
photons~\cite{PhysRevLett.69.1911, BAUER1994471,
PhysRevC.70.031902, PhysRevLett.127.052302, PhysRevLett.98.112001, PhysRevLett.102.222002, 
ALICE:2013wjo, CMS:2012cve, CMS:2018uvs, ATLAS:2020mve,
PhysRevLett.121.132301, 2019134926}.

Theoretically, in 1934 Breit and Wheeler first proposed to study the creation of
electron-positron pairs via the fusion of two real photons,
$\gamma+\gamma\rightarrow{e^{+}+e^{-}}$,
which was difficult to observed in laboratory experiments in that period~\cite{PhysRev.46.1087}.
They calculated the relevant total cross section,
including photon polarization,
and their results were further extended and generalized in Ref.~\cite{TollThesis}.
The above scattering is called the Breit-Wheeler process for the photoproduction of dilepton pairs.
Indeed in 1924 Fermi described the electromagnetic particle production
in terms of an equivalent photon method~\cite{Fermi:1924tc},
which was later improved by Williams~\cite{PhysRev.45.729} and Weizs$\rm\ddot{a}$cker~\cite{ZPhysik1934}.
Within this method, the electromagnetic fields of a moving nucleus can be described as a cloud of quasi-real photons,
and the electromagnetic production cross section of lepton pairs in nuclear collisions can be computed as
\begin{equation}
        \begin{aligned}\label{eq:TradEPA}
                &\sigma^{WW}_{A_{1}A_{2}\rightarrow A_{1}A_{2}l^{+}l^{-}} =
                \int d\omega_{1}d\omega_{2}~n_{1}(\omega_{1})n_{2}(\omega_{2})~\sigma_{\gamma\gamma\rightarrow l^{+}l^{-}},
        \end{aligned}
\end{equation}
in which $n_{1}(\omega_{1})$ and $n_{2}(\omega_{2})$ are the equivalent number of photons with energies $\omega_{1}$
and $\omega_{2}$ from the field of nucleus $A_{1}$ and $A_{2}$, respectively;
$\sigma_{\gamma\gamma\rightarrow l^{+}l^{-}}$ is the elementary two-photon fusion cross section from the Breit-Wheeler process. 
This is known as the equivalent photon approximation (EPA~\cite{osti_4221512, BERTULANI1988299, Vidovic:1992ik, GUCLU19997, BAUR2002359, Bertulani:2005ru, Baur:2007zz, KLUSEKGAWENDA2019339, LI2019576, SuperChi_EPJC19, KLUSEKGAWENDA2021136114, Brandenburg:2022tna, Lin:2022flv, PhysRevC.107.044906}),
and it has been widely employed to interpret
photon-photon collisions in ultra-relativistic heavy-ion collisions~\cite{PhysRevC.70.031902, PhysRevLett.127.052302, PhysRevLett.127.122001, ALICE:2022hvk, PhysRevLett.121.212301, PhysRevC.107.054907}.

Concerning the photon-photon scattering process, $\gamma+\gamma\rightarrow{l^{+}+l^{-}}$,
the momentum of the incoming photons is predominantly along
the beam direction and the relevant transverse momentum is very small,
${k}^{\gamma\gamma}_{\perp}<0.03~{\rm GeV}$~\cite{Bertulani:2005ru, BALTZ20081}.
Thus, according to QED calculation at leading order in $\alpha_{\rm EM}$,
the transverse momentum of the outgoing lepton pairs is also small, $p^{l^{+}l^{-}}_{\perp}\approx0$,
resulting in a nearly back-to-back correlation structure in the azimuthal angle, $|\phi^{l^{+}}-\phi^{l^{-}}|\approx\pi$.
Experimentally, in recent years, the photon-photon scattering processes have been measured
at low $p^{l^{+}l^{-}}_{\perp}$ in Au--Au and Pb--Pb collisions with
nuclear overlap~\cite{PhysRevLett.121.132301, PhysRevLett.121.212301, ALICE:2022hvk, PhysRevC.107.054907},
i.e. their impact parameter $b$ is smaller than twice the nuclear radius, $b<2R$.
It is observed that the back-to-back correlation is broadened and the lepton pair $p^{l^{+}l^{-}}_{\perp}$ increases.

Various theoretical approaches and Monte-Carlo (MC) generators were built to
study the lepton pair photoproduction,
as well as its transverse momentum broadening in heavy-ion collisions.
The widely used model, STARLight~\cite{KLEIN2017258},
is a MC generator with the traditional EPA approach (Eq.~\ref{eq:TradEPA}).
It is designed to simulate a variety of ultra-peripheral collisions (UPC, $b\gtrsim 2R$) without including the photon polarization effects.
SuperChic3~\cite{SuperChi_EPJC19} is a MC generator taking into account the screening effects.
One can run this model without the survival probability, to allow for photon-photon collisions with nuclear overlap.
The photon Wigner formalism~\cite{PhysRevD.101.034015, PhysRevD.102.094013, PhysRevLett.122.132301, KLUSEKGAWENDA2021136114}
allows to provide the calculations of the lepton pair transverse momentum and acoplanarity within a given impact parameter range.
Gamma-UPC~\cite{Shao:2022cly} computes the survival probability of the ions using a parametrized Glauber MC simulation.
One can run this model without requiring survival probability, to allow for photon-photon collisions with nuclear overlap.
It handles final states produced via photon fusion.
All models so far consider only leading-order (LO) QED corrections for the Breit--Wheeler elementary process.
The calculations for the relevant production cross section
are complicated and multi dimensional integration is needed in general to get the numerical results.
Recently, the higher-order (HO) QED effect for the vacuum pair production
was explored by introducing a screening of the Coulomb potential in the
photon propagator~\cite{PhysRevA.61.032103, PhysRevA.64.032106},
which naturally incorporates higher-order corrections~\cite{Zha:2021jhf, Li:2023yjt}.
It is argued that such corrections enhance ($(HO-LO)/LO\approx 15\%$ at maximum) the
transverse momentum broadening of lepton pairs at
energies available at the CERN Large Hadron Collider (LHC)~\cite{Li:2023yjt}.
See Ref.~\cite{Brandenburg:2022tna} (and the references therein) for a comprehensive review of the theoretical models.

We note that, among the various approaches,
the traditional EPA model (Eq.~\ref{eq:TradEPA})
is widely used due to its simplicity~\cite{Vidovic:1992ik, KLEIN2017258, PhysRevC.97.054903}.
Such a model allows us to provide a fairly direct strategy for the calculations
of the Breit-Wheeler process in nuclear collisions.
However, as pointed out in Ref.~\cite{Brandenburg:2022tna},
the traditional approach has difficulty in computing
the kinematic distributions of single lepton,
which are challenged by the recent measurements performed at
energies available at the BNL Relativistic Heavy Ion Collider (RHIC) and at the LHC.
In this work, we try to overcome this issue by utilizing a Monte-Carlo-based strategy (see Sec.~\ref{subsec:MCsetup}).
The differential cross section for the photon-photon scattering process, $\gamma+\gamma\rightarrow{l^{+}+l^{-}}$,
is calculated via the traditional EPA (Eqs.~\ref{eq:DiffSigma_OmegaB} and~\ref{eq:DiffSigma_WYB}).
The relevant single photon transverse momentum is sampled according to
the differential form of the one-photon distribution function (Eq.~\ref{eq:C9}),
which is obtained by integrating over the full transverse plane
perpendicular to the direction of the moving nucleus.
After including the fusion process for $\gamma+\gamma\rightarrow{l^{+}+l^{-}}$,
a hybrid model is developed to provide the impact parameter dependent calculations for events with nuclear overlap.
The latest experimental data, such as the transverse momentum broadening
and the invariant mass spectrum, will be used to examine the relevant theoretical calculations.

In summary, the different models/MCs of the Breit--Wheeler process mentioned above
contain different corrections to account for the finite (low) virtuality of the
incoming photon fluxes (and subsequent propagated $k_{\perp}$ to the outgoing $l^{\pm}$),
and in some cases too for the incoming polarization of the photons
(and subsequent azimuthal modulation of the final-state dielectrons),
and have implemented different methods to take into account the survival probability of the ions.
This work is just based on an alternative approach to model the Breit--Wheeler
process in photon-photon collisions with nuclear overlap.

The paper is organized as follows:
Sec.~\ref{sec:Method} is dedicated to the description of the theoretical framework,
including the calculation of the electromagnetic field,
the equivalent photon spectra and the Monte-Carlo-based setup.
Sec.~\ref{sec:result} shows the results of transverse momentum broadening and the invariant mass spectrum.
A summary section then follows.

\section{Theoretical method and configuration}\label{sec:Method}

\subsection{The equivalent photon approximation}\label{subsec:EPAmodel}
To calculate the energy flux of the classical equivalent photons,
we first determine the electromagnetic field of a charge distribution moving at high velocity ($v\approx c$),
and then calculate the single photon distributions.
Here we just show the final results, and the detailed aspects are relegated to the Appendix-\ref{app:appendix}.

In the observer's frame, we assume a nucleus moves with constant velocity $v$ on
straight line along the $z$-axis (i.e. longitudinal direction),
and being located by the displacement $\vec{b}$ in the $xy$-plane (i.e. transverse plane).
The relevant potentials $A(x)=(t,\vec{x}_{\perp}-\vec{b},z)$
of the electromagnetic waves are determined by d'Alembert's equation
\begin{equation}
        \begin{aligned}\label{eq:dAlbertEq}
                &\partial_{\mu}\partial^{\mu}A^{\nu}(x) = J^{\nu}(x)
        \end{aligned}
\end{equation}
in the Lorentz gauge
\begin{equation}
        \begin{aligned}\label{eq:LorentzGauge}
        \partial_{\nu}A^{\nu}(x)=0.
        \end{aligned}
\end{equation}
We perform the Fourier transformation of Eq.~\ref{eq:dAlbertEq}, yielding (Eq.~\ref{eq:A8})
\begin{equation}
        \begin{aligned}\label{eq:4PotentialInMom}
                A^{\nu}(k) = -\frac{1}{k^{2}}J^{\nu}(k) = 2\pi Ze~ \delta(ku) \frac{\mathcal{F}(\sqrt{{-k^{2}}})}{-k^{2}} u^{\nu} e^{i\vec{k}\cdot \vec{b}},
        \end{aligned}
\end{equation}
with $u=\gamma(1,\vec{v})=\gamma(1,\vec{0}_{\perp},v)$ and $k=(\omega,\vec{k})=(\omega,\vec{k}_{\perp},\omega/v)$.
$Z$ is the charge number of the nucleus, and $\mathcal{F}$ is the corresponding form factor
\begin{equation}\label{eq:FormFactor}
        \mathcal{F}(q^{2}) \equiv \int^{\infty}_{-\infty} d^{3}\vec{r} e^{-i\vec{q}\cdot\vec{r}} \rho(\vec{r})
\end{equation}
where the four-momentum transfer reads $q^{2}\equiv -k^{2}\doteq \vec{k}^{\;2}_{\perp} + (\omega/\gamma)^{2}$ in the ultra-relativistic limit $v\approx c=1$.
The spatial charge distribution $\rho$ is quantified by the Woods-Saxon distribution~\cite{PhysRevC.97.054910}
\begin{equation}\label{eq:WoodsSaxon}
        \rho(r)=\rho_{0} \frac{1+\Omega\cdot(r/R)^{2}} {1+\exp[(r-R)/a]},
\end{equation}
where, $\rho_{0}$ is the normalization factor so that $\int d^{3}\vec{r}\rho(\vec{r})=1$,
$r=|\vec{r}|$ is the distance with respect to the nucleus center,
$R$ is the nucleus radius, $a$ is the skin depth,
and $\Omega$ corresponds to deviations from a spherical shape for a given nucleus.
The employed values of these three parameters in this work are summarized in Tab.~\ref{tab:WodsSax_Parameter}.
\begin{table}[!htbp]
\centering
\begin{tabular}{c|c|c|c}
\hline                                                                                                                                                                       
\multicolumn{1}{l}{\multirow{1}{*}{\centering Nucleus\qquad}}
 & \multicolumn{1}{l}{\multirow{1}{*}{\centering $R$ (fm)\qquad}}
 & \multicolumn{1}{l}{\multirow{1}{*}{\centering $a$ (fm)\qquad}}
 & \multicolumn{1}{l}{\multirow{1}{*}{\centering $\Omega$}}
  \\
\hline
\multicolumn{1}{l}{\centering $^{197}$Au}
 & \multicolumn{1}{l}{\centering 6.380}
 & \multicolumn{1}{l}{\centering 0.535}
 & \multicolumn{1}{l}{\centering 0}
 \\
\multicolumn{1}{l}{\centering $^{208}$Pb}
 & \multicolumn{1}{l}{\centering 6.624}
 & \multicolumn{1}{l}{\centering 0.549}
 & \multicolumn{1}{l}{\centering 0}
 \\
\hline
\end{tabular}
\caption{Nuclear density parameter for the charge density distributions of Au and Pb.
Results adopted from Ref.~\cite{PhysRevC.97.054910}.}
\label{tab:WodsSax_Parameter}
\end{table}

The resulting form factor in Eq.~\ref{eq:FormFactor} can be rewritten as~\cite{zbMATH03059048}
\begin{equation}
        \begin{aligned}\label{eq:FormFactor_Simple}
                \mathcal{F}(q^{2}) =& \frac{4\pi^{2}\rho_{0}a^{3}}{(qa)^{2}~sinh^{2}(\pi qa)} \\
                &{\bigr[} \pi qa~cosh(\pi qa)~sin(qR) - qR~cos(qR)~sinh(\pi qa) {\bigr]} \\
                &+8\pi\rho_{0}a^{3} \Sigma^{\infty}_{j=1} (-1)^{j-1} \frac{je^{-jR/a}}{[j^{2}+(qa)^{2}]^{2}},
        \end{aligned}
\end{equation}
in which the term on the third line is expected to be much smaller than the others.
We have checked its validity and therefore neglected it in this analysis.
A similar approximation is widely employed in the literature~\cite{2016EPJC76.428S}.


Here we treat the electromagnetic field of the two nuclei classically by
assuming both nuclei move with constant velocity on straight lines being separated by
the impact parameter $\vec{b}$.
The differential cross section for the photon-photon scattering process,
\begin{widetext}
$\gamma+\gamma\rightarrow l^{+}+l^{-}$, reads~\cite{Vidovic:1992ik}
\begin{equation}
        \begin{aligned}\label{eq:DiffSigma_OmegaB}
                \frac{d^{4}\sigma}{d\omega_{1}d\omega_{2}d^{2}\vec{b}} =
                ~n_{s}(\omega_{1},\omega_{2},\vec{b})\sigma_{s}(\omega_{1},\omega_{2})
                  + ~n_{ps}(\omega_{1},\omega_{2},\vec{b})\sigma_{ps}(\omega_{1},\omega_{2}),
        \end{aligned}
\end{equation}
in which $n_{s}(\omega_{1},\omega_{2},\vec{b})$ and $n_{ps}(\omega_{1},\omega_{2},\vec{b})$
are the scalar and pseudo-scalar two-photon distribution functions, respectively.
Such distribution functions can be expressed in terms of the electromagnetic fields,
\begin{equation}
        \begin{aligned}\label{eq:TwoPhoton_scaler}
                n_{s}(\omega_{1},\omega_{2},\vec{b})
                &= \frac{1}{\pi^{2}\omega_{1}\omega_{2}} \int d^{2}\vec{x}_{\perp} {\biggr[} \vec{E}_{1\perp}(\omega_{1},\vec{x}_{\perp}-\vec{b},z)\cdot
                \vec{E}_{2\perp}(\omega_{2},\vec{x}_{\perp},z) {\biggr]}^{2} \\
                &\stackrel{(\ref{eq:C3})}{=} \int d^{2}\vec{x}_{\perp} n(\omega_{1},\vec{x}_{\perp}-\vec{b})~n(\omega_{2},\vec{x}_{\perp}) {\biggr[}
                \frac{(\vec{x}_{\perp}-\vec{b})\cdot\vec{x}_{\perp}}{|\vec{x}_{\perp}-\vec{b}|\cdot|\vec{x}_{\perp}|} {\biggr]}^{2},
        \end{aligned}
\end{equation}
\begin{equation}
        \begin{aligned}\label{eq:TwoPhoton_psudoscaler}
                n_{ps}(\omega_{1},\omega_{2},\vec{b})
                &= \frac{1}{\pi^{2}\omega_{1}\omega_{2}} \int d^{2}\vec{x}_{\perp} {\biggr[} \vec{E}_{1\perp}(\omega_{1},\vec{x}_{\perp}-\vec{b},z)\times
                \vec{E}_{2\perp}(\omega_{2},\vec{x}_{\perp},z) {\biggr]}^{2} \\
                &\stackrel{(\ref{eq:C3})}{=} \int d^{2}\vec{x}_{\perp} n(\omega_{1},\vec{x}_{\perp}-\vec{b})~n(\omega_{2},\vec{x}_{\perp}) {\biggr[}
                \frac{(\vec{x}_{\perp}-\vec{b})\times\vec{x}_{\perp}}{|\vec{x}_{\perp}-\vec{b}|\cdot|\vec{x}_{\perp}|} {\biggr]}^{2},
        \end{aligned}
\end{equation}
in the ultra-relativistic limit $v\approx c=1$.
$n(\omega,\vec{x}_{\perp})$ is the one-photon distribution function, as given in Eq.~\ref{eq:C6}.
The scalar part $n_{s}$ and the pseudo-scalar part $n_{ps}$
correspond to the electric fields that are parallel, $\vec{E}_{1}\parallel\vec{E}_{2}$,
and perpendicular, $\vec{E}_{1}\perp\vec{E}_{2}$, respectively.

The elementary two-photon fusion cross section in Eq.~\ref{eq:DiffSigma_OmegaB}
is expressed as~\cite{osti_42215121}
\begin{equation}
        \begin{aligned}\label{eq:Sigma_scaler}
                \sigma^{\gamma\gamma\rightarrow l^{+}l^{-}}_{s}&=\frac{4\pi\alpha^{2}_{EM}}{s} {\biggr[} (1+\frac{4m^{2}_{l}}{s}-\frac{12m^{4}_{l}}{s^{2}}) 2ln(\frac{\sqrt{s}}{2m_{l}}+ \sqrt{\frac{s}{4m^{2}_{l}}-1})
                -(1+\frac{6m^{2}_{l}}{s})\sqrt{1-\frac{4m^{2}_{l}}{s}} {\biggr]} \Theta(s-4m^{2}_{l}) \\
        \end{aligned}
\end{equation}

\begin{equation}
        \begin{aligned}\label{eq:Sigma_psudoscaler}
                \sigma^{\gamma\gamma\rightarrow l^{+}l^{-}}_{ps}&=\frac{4\pi\alpha^{2}_{EM}}{s} {\biggr[} (1+\frac{4m^{2}_{l}}{s}-\frac{4m^{4}_{l}}{s^{2}}) 2ln(\frac{\sqrt{s}}{2m_{l}}+ \sqrt{\frac{s}{4m^{2}_{l}}-1})
                -(1+\frac{2m^{2}_{l}}{s})\sqrt{1-\frac{4m^{2}_{l}}{s}} {\biggr]} \Theta(s-4m^{2}_{l})
        \end{aligned}
\end{equation}
\end{widetext}
at the leading order in $\alpha_{EM}$, where, $s=4\omega_{1}\omega_{2}$ indicates the center-of-mass squared-energy
and $m_{l}$ denotes the mass of single lepton $l$.
The step function $\Theta(s-4m^{2}_{l})$ guarantees that the center-of-mass energy of the two photons is
no smaller than twice the lepton mass.

The angular profile of the produced lepton pairs reads
\begin{equation}
        \begin{aligned}\label{eq:angular}
G(\theta)=2+4\biggr(1-\frac{4m_{l}^{2}}{W^{2}}\biggr)\frac{(1-\frac{4m_{l}^{2}}{W^{2}})sin^{2}\theta\;cos^{2}\theta+\frac{4m_{l}^{2}}{W^{2}}}{\bigr[ 1-(1-\frac{4m_{l}^{2}}{W^{2}})cos^{2}\theta\bigr]^{2}}
      \end{aligned}
\end{equation}
where, $\theta$ is the angle between the beam direction and one of the leptons in the local rest frame of the lepton pair.
The result is adopted from Refs.~\cite{Stanley71PRD, KLEIN2017258} by neglecting the
effect of the photon momentum on the angular distribution.

It is convenient to show the results in terms of the invariant mass $M^{\gamma\gamma}$
and rapidity $Y^{\gamma\gamma}$ of the $\gamma\gamma$ system,
so that we implement further calculations by performing a simple change of variables.
The center-of-mass energy reads
\begin{equation}
        \begin{aligned}\label{eq:InvMas}
                W\equiv M^{\gamma\gamma}&=\sqrt{{(E^{\gamma\gamma})}^{2}-(\vec{p}^{\;\gamma\gamma})^{2}} \doteq \sqrt{4\omega_{1}\omega_{2}}
        \end{aligned}
\end{equation}
and the rapidity is
\begin{equation}
        \begin{aligned}\label{eq:Rap}
                Y\equiv Y^{\gamma\gamma}= \frac{1}{2}ln \frac{E^{\gamma\gamma}+p^{\gamma\gamma}_{z}}{E^{\gamma\gamma}-p^{\gamma\gamma}_{z}}
                = \frac{1}{2}ln \frac{\omega_{1}}{\omega_{2}}.
        \end{aligned}
\end{equation}
Therefore, the single photon energy can be determined by
\begin{equation}
        \begin{aligned}\label{eq:W1W2}
                &\omega_{1}=\frac{W}{2} ~ e^{Y} \qquad \omega_{2}=\frac{W}{2} ~ e^{-Y}
        \end{aligned}
\end{equation}
and the right hand side of Eq.~\ref{eq:DiffSigma_OmegaB} can be rewritten as
\begin{equation}
        \begin{aligned}\label{eq:DiffSigma_WYB}
                \frac{d^{4}\sigma}{d\omega_{1}d\omega_{2}d^{2}\vec{b}}=\frac{\partial(W,Y)}{\partial(\omega_{1},\omega_{2})}    \frac{d^{4}\sigma}{dWdYd^{2}\vec{b}} =\frac{2}{W} \frac{d^{4}\sigma}{dWdYd^{2}\vec{b}}.
        \end{aligned}
\end{equation}

\subsection{Monte Carlo based setup}\label{subsec:MCsetup}
It is realized that, in Eqs.~\ref{eq:DiffSigma_OmegaB} and \ref{eq:DiffSigma_WYB},
the theoretical calculations explicitly integrated over
the momentum and (pseudo-)rapidity of single lepton,
which are restricted in the available measurements.
To better perform the comparison with data,
we propose a Monte Carlo based strategy to overcome this issue.

In this sub-section, we describe the numerical framework utilized for
the photoproduction of lepton pairs, $\gamma\gamma\rightarrow l^{+}l^{-}$,
in ultra-relativistic heavy-ion collisions.
Generally, in the center-of-mass (CM) frame of the $\gamma\gamma$ system,
the energy of each photon can be obtained at a given rapidity,
while its transverse momentum is sampled according to the single photon energy spectrum.
Then, the fusion process for $\gamma\gamma\rightarrow l^{+}l^{-}$ is
performed by employing a Monte-Carlo-based setup.
Finally, we can boost the above results from CM to the laboratory (LAB).
The steps of this numerical procedure are
\begin{enumerate}
\item[(1)] Calculate the differential production cross section
$\frac{{d^{4}\sigma}}{{dWdYd^{2}\vec{b}}}$ (Eq.~\ref{eq:DiffSigma_WYB}) at a given range of $(W, Y, b)$;
note that the varying ranges for $W$ and $Y$ are determined by the corresponding measurements,
while $b$ for a desired centrality class is given in Ref.~\cite{PhysRevC.97.054910}.

\item[(2)] Sample a set of $\gamma\gamma$ pairs according to the profile of the above spectra via Monte-Carlo,
and then initialize their four-momentum, in the CM frame of each pair,
at a given point $(W, Y, b)$
        \begin{itemize}
        \item[$\bullet$] the total energy of the $\gamma\gamma$ system: $E^{Tot}=\omega_{1}+\omega_{2}$,
        with the energy of single photon $i$, $\omega_{i}$, given by Eq.~\ref{eq:W1W2};

        \item[$\bullet$] the total transverse momentum: $\vec{P}^{Tot}_{\perp}=\vec{k}_{1,\perp}+\vec{k}_{2,\perp}$,
        with transverse momentum of photon $i$, $\vec{k}_{i\perp}$, sampled according to
        the single photon energy spectrum (Eq.~\ref{eq:C9}) at a given energy $\omega_{i}$;
        the azimuthal angle of $\vec{P}^{Tot}_{\perp}$ is assumed to be uniformly distributed in the transverse plane,
        and it will be constrained by the kinematics cuts on the decaying lepton pairs;
	
        \item[$\bullet$] the total longitudinal momentum: \newline $P^{Tot}_{z}=\sqrt{(W)^{2}+(\vec{P}^{Tot}_{\perp})^{2}}\cdot sinh(Y)$;
        \end{itemize}

\item[(3)] Perform the fusion process $\gamma\gamma\rightarrow l^{+}l^{-}$
        \begin{itemize}
        \item[$\bullet$] calculate the four-momentum for the $i$-th lepton $(E_{i}$, $\vec{p}_{i,\perp}$, $p_{i,z})$
                \begin{itemize}
                \item initialize its three-momentum according to the two-body decay kinematics
                $|\vec{p}_{i}|=\sqrt{W^{2}-4m^{2}_{l}}/2$ in the CM of $\gamma\gamma$ system,
		as well as the energy $E_{i}=\sqrt{(\vec{p}_{i})^{2}+m^{2}_{l}}$;

                \item sample the transverse momentum $\vec{p}_{i,\perp}=(|\vec{p}_{i}|sin\theta cos\phi, |\vec{p}_{i}|sin\theta sin\phi)$
		and the longitudinal momentum $p_{i,z}=|\vec{p}_{i}|cos\theta$ by assuming that
		the polar angle ($\theta$) profile is described by Eq.~\ref{eq:angular},
		while the azimuthal angle ($\phi$) is uniformly distributed;
		they will be further constrained by the additional kinematics cuts, which are imposed in the corresponding final observable;
                \end{itemize}

        \item[$\bullet$] boost the above results from the CM of $\gamma\gamma$ system to the LAB
        according to $\vec{\beta}=(\beta_{x},\beta_{y},\beta_{z})$, where $\beta_{i}=-P^{Tot}_{i}/E^{Tot}$ ($i=x,y,z$);
        note that $\vec{P}^{Tot}$ and $E^{Tot}$ are already determined in the previous step;

        \item[$\bullet$] boost further the obtained results from the LAB to CM for the asymmetry colliding system $A_{1}A_{2}$,
        according to $\vec{\beta}=(0, 0, \beta_{z})$, where $\beta_{z}=tanh(Y_{shift})$ with
        the rapidity shift $Y_{shift}=\frac{1}{2}ln{\bigr(} \frac{Z_{1}}{A_{1}} \frac{A_{2}}{Z_{2}} {\bigr)}$ and
        $Z_{i}$ and $A_{i}$ are the charge number and mass number of nucleus $i$, respectively;
        \end{itemize}

\item[(4)] Collect the four-momentum of lepton pairs $l^{+}l^{-}$ and single leptons within the desired acceptance,
which can be used to calculate observables such as
$\frac{dN}{dW}$, $\frac{dN}{dp^{2}_{\perp}}$, and $\sqrt{<(p_{\perp})^{2}>}$,
and then compared to the relevant measurements within the same acceptance.

\end{enumerate}

\section{Numerical results}\label{sec:result}

\begin{figure}[!htbp]
\begin{center}
\includegraphics[width=0.40\textwidth]{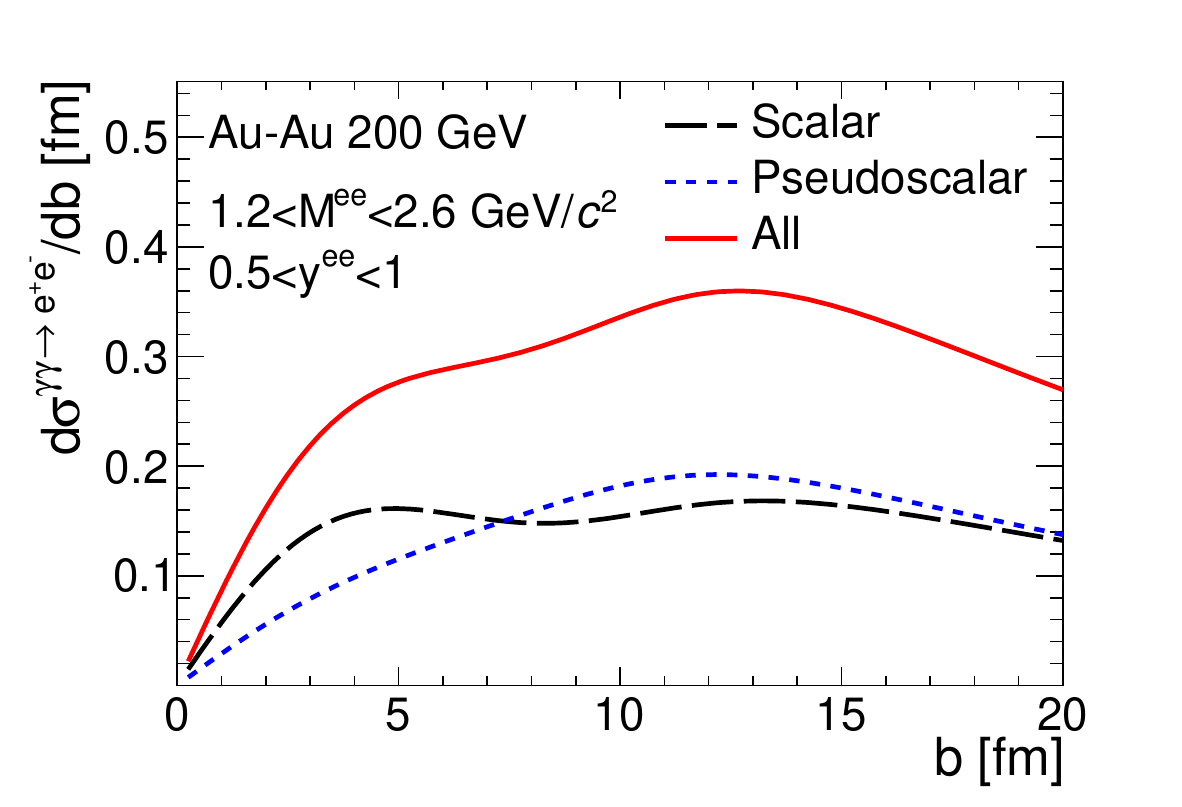}
\caption{(Color online) Impact parameter dependent differential cross section (solid red curve)
for the electromagnetic production of an electron-positron pair
with scalar (dashed black curve) and pseudoscalar contribution (dotted blue curve),
in Au--Au collisions at $\snn=200$ GeV.
See legend for details.}
\label{fig:RHIC_dSigmadB}
\end{center}
\end{figure}

Figure~\ref{fig:RHIC_dSigmadB} depicts the impact parameter dependent cross section
for the electromagnetic production of an electron-positron pair in Au--Au collisions at $\snn=200$ GeV.
The scalar and pseudoscalar contributions (Eq.~\ref{eq:DiffSigma_OmegaB})
are displayed as dashed black and dotted blue curves, respectively,
while the combined result is presented as a solid red curve.
We observe that, for the scalar component, a dip structure lies at $b\approx 9$ fm,
while for the pseudoscalar component, a maximum is found at $b\approx 12$ fm.

\begin{figure}[!htbp]
\begin{center}
\includegraphics[width=.40\textwidth]{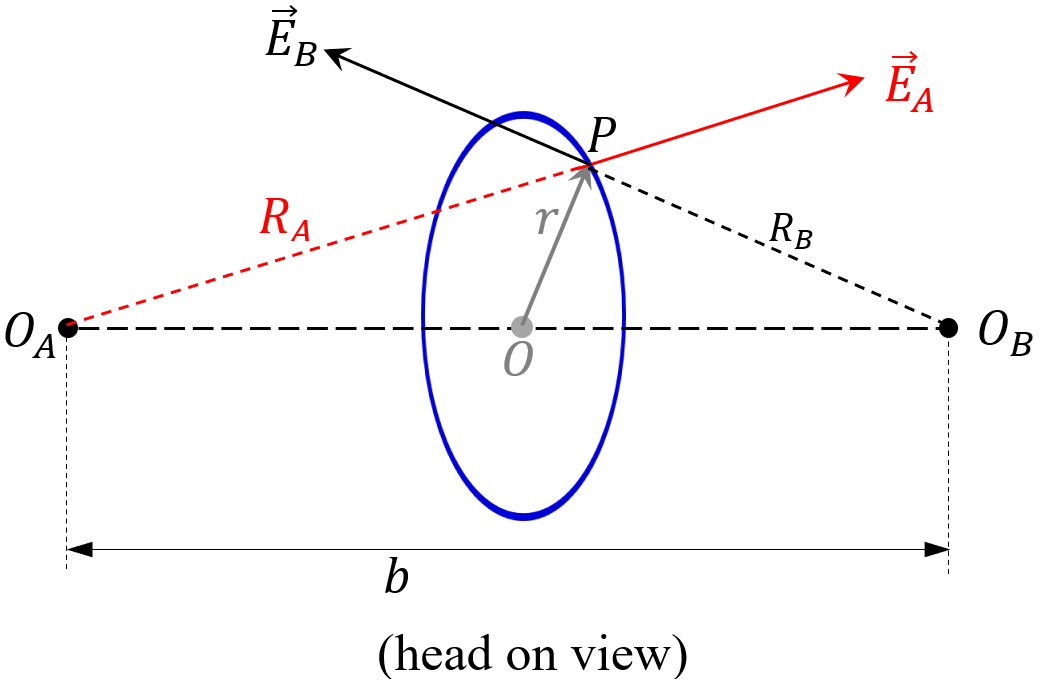}
\caption{Illustration of the electric field produced by two colliding heavy ions.}
\label{fig:sketch}
\end{center}
\end{figure}
These behaviors may be induced by the two overlapping electromagnetic fields:
(1) the electron-positron pairs are more often produced in the vicinity of the nuclear surfaces,
where the strongest electric field densities of the two colliding heavy ions exist~\cite{PhysRevC.85.044907};
(2) for the two colliding heavy ions $AB$ (see the illustration in Fig.~\ref{fig:sketch}),
the overlap nuclear surface is shown as the blue circle,
with an arbitrary point $P$ on top of it to display the point of interest ($b\le R_{A}+R_{B}$);
the constellation of the electric fields are mostly perpendicular, $\vec{E}_{\rm A}\perp\vec{E}_{\rm B}$,
when $b=\sqrt{R^{2}_{\rm A}+R^{2}_{\rm B}}$;
similarly, they are mostly parallel, $\vec{E}_{\rm A}\parallel\vec{E}_{\rm B}$, when $b=R_{\rm A}+R_{\rm B}$.
As mentioned in Eqs.~\ref{eq:TwoPhoton_scaler} and \ref{eq:TwoPhoton_psudoscaler},
the electric fields of the two colliding heavy ions have to be parallel and perpendicular
to produce the scalar and pseudoscalar two-photon fusion cross sections, respectively.
Thus, for the scalar contribution, one can expect that a dip structure exists around $b=\sqrt{2}R_{\rm Au}\approx 9$ fm,
while, for the pseudoscalar part, its maximum falls into the dip of the double hump structure of the scalar cross section,
i.e., the maximum lies below $b=2R_{\rm Au}\approx13$ fm in Au--Au collisions.
Similar results can be found in Refs.~\cite{Vidovic:1992ik, KRAUSS1997503}.

\begin{figure}[!htbp]
\begin{center}
\includegraphics[width=.40\textwidth]{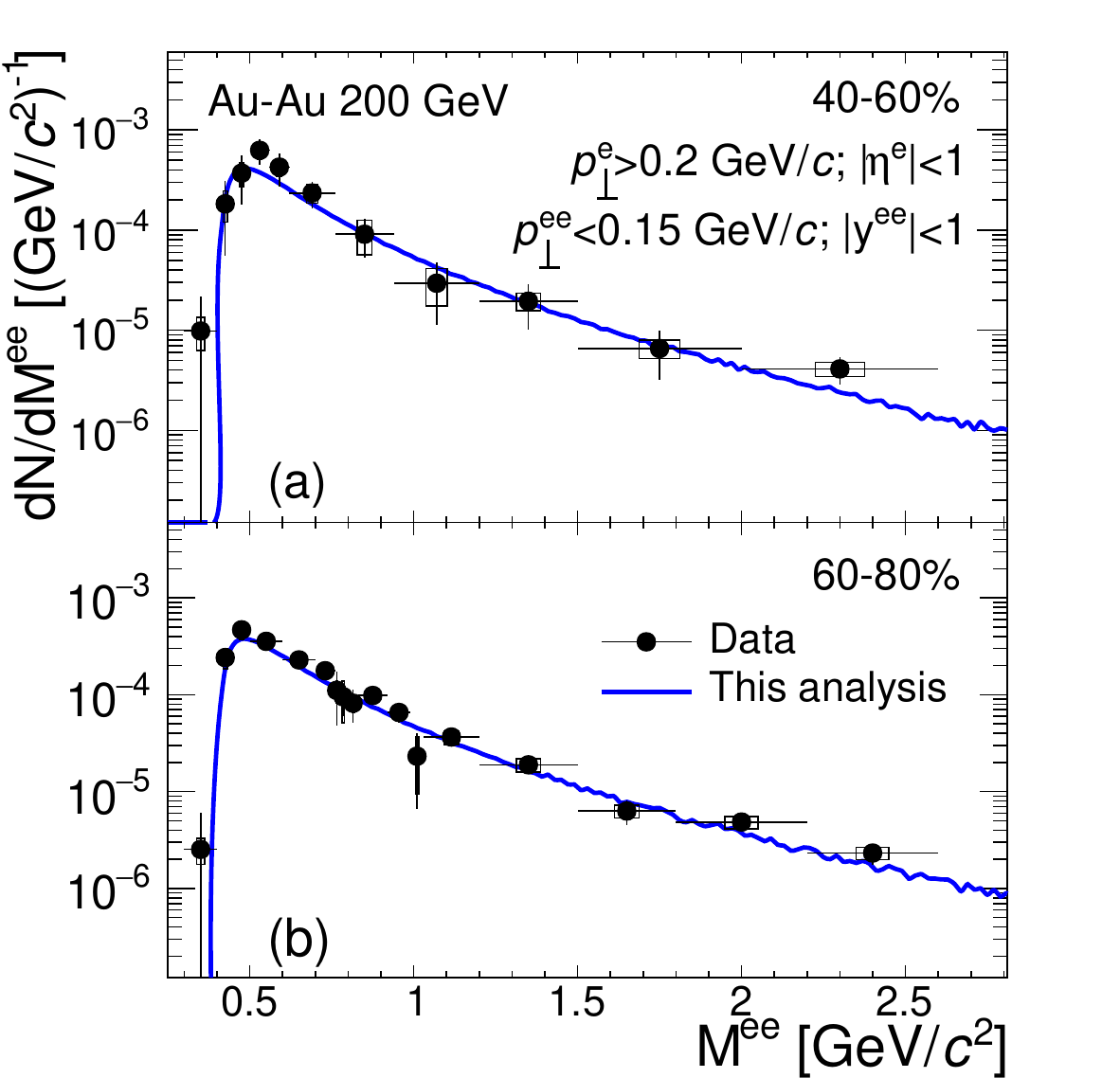}
\caption{(Color online) The differential yield with respect to dielectron invariant mass in mid-central (panel-a) and peripheral (panel-b) collisions,
indicating as the centrality percentile $40-60\%$ and $60-80\%$, respectively,
Au--Au collisions at $\snn=200$ GeV.
The results are filtered within the selected acceptance.
The relevant data (solid black point~\cite{PhysRevLett.121.132301}) are shown for comparison.
See legend and text for details.}
\label{fig:RHIC_dNdM}
\end{center}
\end{figure}
\begin{figure}[!htbp]
\begin{center}
\includegraphics[width=.40\textwidth]{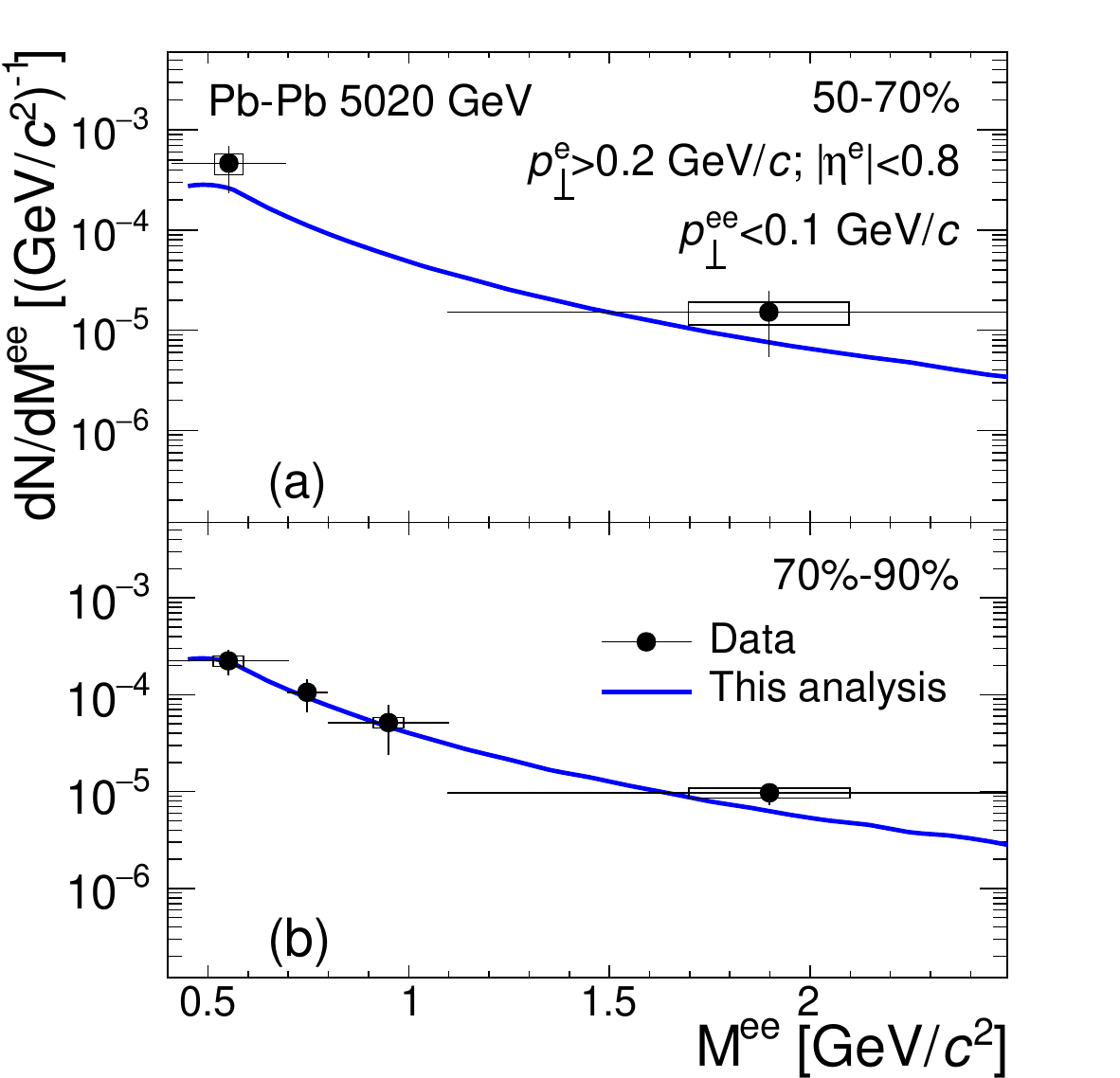}
\caption{Same as Fig.~\ref{fig:RHIC_dNdM} but in $50-70\%$ (panel-a) and $70-90\%$ (panel-b) Pb--Pb collisions at $\snn=5.02$ TeV.
The relevant data (solid black point~\cite{ALICE:2022hvk}) are shown for comparison.}
\label{fig:ALICE_dNdM}
\end{center}
\end{figure}

Figure~\ref{fig:RHIC_dNdM} shows the normalized differential yield as a function of the
invariant mass of $e^{+}e^{-}$ in mid-central ($40-60\%$, panel-a) and peripheral ($60-80\%$, panel-b) Au--Au collisions at $\snn=200$ GeV.
The results (curves) are restricted within the kinematics region
($\pt^{e}>0.2~{\rm GeV}/{\it c}$, $|\eta^{e}|<1$, $\pt^{ee}<0.15~{\rm GeV}/{\it c}$ and $|y^{ee}|<1$),
as applied in the STAR measurements, for direct comparisons with the experimental data (points).
Due to the conservation laws of energy and momentum,
the spectra are zero when the invariant mass is less than twice the transverse momentum of the single electron,
$M^{ee}\lesssim0.4~{\rm GeV}/c^{2}$.
It peaks at $M^{ee}\approx 0.5~{\rm GeV}/c^{2}$ and then decreases exponentially towards larger $M^{ee}$.
It can be seen that the model calculations provide a very good description
of the measured $M^{ee}$-dependent spectra data in both $40-60\%$ (panel-a) and $60-80\%$ (panel-b) centrality classes.
Similar behavior was found for Pb--Pb collisions at $\snn=5.02$ TeV
in different centrality classes, as shown in Fig.~\ref{fig:ALICE_dNdM}.

\begin{figure}[!htbp]
\begin{center}
\includegraphics[width=.40\textwidth]{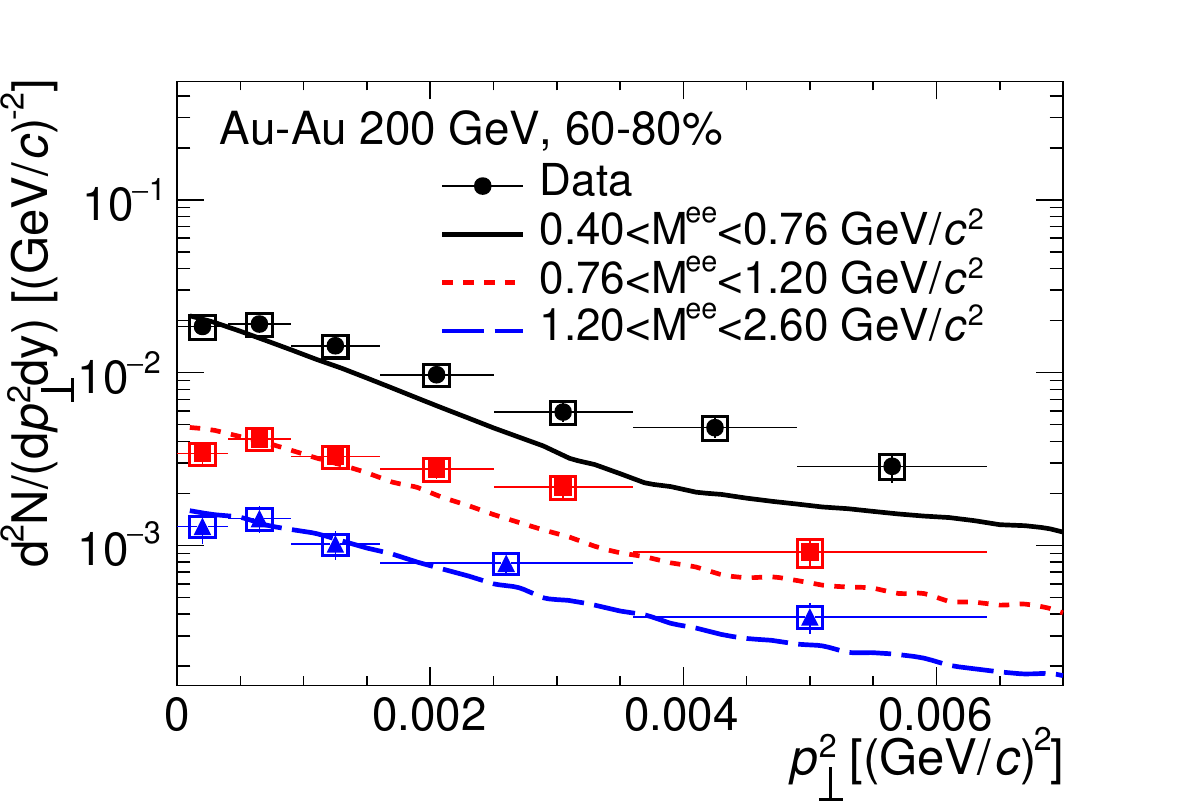}
\caption{(Color online) The dielectron pair-$\pt^{2}$ spectrum within the mass regions
$0.4<M^{ee}<0.76~{\rm GeV}/c^{2}$ (solid black curve), $0.76<M^{ee}<1.2~{\rm GeV}/c^{2}$ (dotted red curve)
and $1.2<M^{ee}<2.6~{\rm GeV}/c^{2}$ (dashed blue curve) in $60-80\%$ Au--Au collisions at $\snn=200$ GeV.
The comparisons with available STAR measurements~\cite{PhysRevLett.121.132301} are plotted as well.}
\label{fig:RHIC_dNdpT2}
\end{center}
\end{figure}

In Fig.~\ref{fig:RHIC_dNdpT2}, the differential electron-positron pair $\pt^{2}$ spectrum
$d^{2}N/(d\pt^{2}dy)$ is calculated within three invariant mass regions
in $60-80\%$ Au--Au collisions at $\snn=200$ GeV.
See the legend for details.
The numerical calculations (curves) are filtered with the STAR
acceptance ($\pt^{e}>0.2~{\rm GeV}/{\it c}$, $|\eta^{e}|<1$ and $|y^{ee}|<1$),
allowing direct comparison with the experimental measurements (points).
Within the experimental uncertainties,
the measured $\pt^{2}$ dependencies are compatible with the model predictions,
while a slightly larger discrepancy observed
in the range $\pt^{2}\gtrsim 0.002~({\rm GeV}/\it{c})^{\rm 2}$.

\begin{figure}[!htbp]
\begin{center}
\includegraphics[width=.40\textwidth]{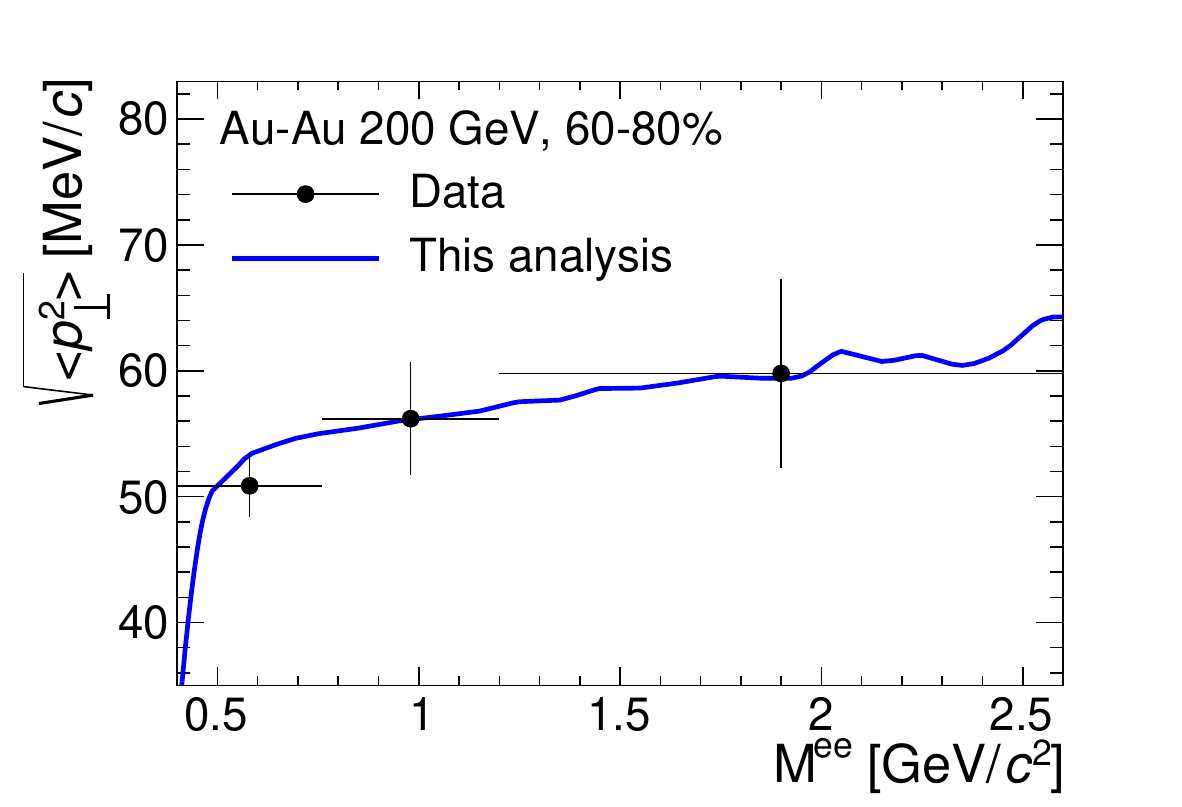}
\caption{The dielectron pair-$\sqrt{<\pt^{2}>}$ distribution as a function of invariant mass
in $60-80\%$ Au--Au collisions at $\snn=200$ GeV.
The measurements~\cite{PhysRevLett.121.132301} are also shown for comparisons.}
\label{fig:RHIC_AvgPtvsM}
\end{center}
\end{figure}

Figure~\ref{fig:RHIC_AvgPtvsM} presents the average transverse momentum
squared $\sqrt{<\pt^{2}>}$ of electron-positron pairs as a function of $M^{ee}$
in $60-80\%$ Au--Au collisions at $\snn=200$ GeV.
It is observed that $\sqrt{<\pt^{2}>}$ is more significant at larger $M^{ee}$,
where the electron-positron pairs are generated predominantly
in the vicinity of the stronger electromagnetic field,
which, in turn, creates larger $\pt$.
The available STAR measurements~\cite{PhysRevLett.121.132301} are displayed as well for comparison.
Both the shape and the magnitude of the measured data can be described reasonably well by this calculation,
thereby reducing the available space for the additional effects,
such as the higher-order QED corrections and the hot medium in collisions with nuclear overlap~\cite{Li:2023yjt}.
This data-to-model comparison can be improved with future high-precision measurements,
which are important to quantify the spread of dilepton $\pt$ distributions, in particular in more central collisions.
Table~\ref{tab:AvgPtSqu_Cmp} summarizes the $\sqrt{<\pt^{2}>}$ measured in three different mass regions
in Pb--Pb collisions at $\snn=5.02~{\rm TeV}$~\cite{ALICE:2022hvk}.
Within the experimental uncertainties they are found to be in agreement with our results.
Note that the model predictions based on the lowest-order QED calculations (``QED''~\cite{Zha:2018tlq, Brandenburg:2021lnj}),
the Wigner formalism (``Wigner''~\cite{KLUSEKGAWENDA2021136114}),
and the vanishing impact parameter effects (``STARLight''~\cite{KLEIN2017258, PhysRevC.97.054903})
are shown as well for comparisons.
All the results from theory present an increasing behavior,
while the data are not yet precise enough to conclude such mass dependence of $\sqrt{<\pt^{2}>}$~\cite{ALICE:2022hvk}.
\begin{table*}[!htbp]
\centering
\begin{tabular}{c|c|c|c|c|c|}
\hline
\multicolumn{1}{c}{\multirow{1}{*}{Mass region}}
 & \multicolumn{1}{c}{\multirow{1}{*}{\centering ALICE Data~\cite{ALICE:2022hvk}}}
 & \multicolumn{1}{c}{\multirow{1}{*}{\centering QED~\cite{Zha:2018tlq, Brandenburg:2021lnj}}}
 & \multicolumn{1}{c}{\multirow{1}{*}{\centering Wigner~\cite{KLUSEKGAWENDA2021136114}}}
 & \multicolumn{1}{c}{\multirow{1}{*}{\centering STARLight~\cite{KLEIN2017258, PhysRevC.97.054903}}}
 & \multicolumn{1}{c}{\multirow{1}{*}{\centering This work}}
  \\
\multicolumn{1}{c}{\centering ($\rm GeV/{\it{c}}^2$)}
 & \multicolumn{1}{c}{\centering \quad ($\rm MeV/{\it{c}})$ \quad}
 & \multicolumn{1}{c}{\centering \quad ($\rm MeV/{\it{c}})$ \quad}
 & \multicolumn{1}{c}{\centering \quad ($\rm MeV/{\it{c}})$ \quad}
 & \multicolumn{1}{c}{\centering \quad ($\rm MeV/{\it{c}})$ \quad}
 & \multicolumn{1}{c}{\centering \quad ($\rm MeV/{\it{c}}$)}
 \\
\hline
\multicolumn{1}{c}{\centering $0.4 \le M^{ee}\le 0.7$}
 & \multicolumn{1}{c}{\centering $\qquad 44\pm 28(stat.)\pm 6(syst.) \qquad$}
 & \multicolumn{1}{c}{\centering 44}
 & \multicolumn{1}{c}{\centering 45}
 & \multicolumn{1}{c}{\centering 30}
 & \multicolumn{1}{c}{\centering 39}
 \\
\hline
\multicolumn{1}{c}{\centering $0.7 \le M^{ee}\le 1.1$}
 & \multicolumn{1}{c}{\centering $\qquad 45\pm 36(stat.)\pm 8(syst.) \qquad$}
 & \multicolumn{1}{c}{\centering 48}
 & \multicolumn{1}{c}{\centering 48}
 & \multicolumn{1}{c}{\centering 38}
 & \multicolumn{1}{c}{\centering 43}
 \\
 \hline
\multicolumn{1}{c}{\centering $1.1 \le M^{ee}\le 2.7$}
 & \multicolumn{1}{c}{\centering $\qquad 69\pm 36(stat.)\pm 8(syst.) \qquad$}
 & \multicolumn{1}{c}{\centering 50}
 & \multicolumn{1}{c}{\centering 50}
 & \multicolumn{1}{c}{\centering 42}
 & \multicolumn{1}{c}{\centering 45}
 \\
\hline
\end{tabular}
\caption{Summary of the different models for $\sqrt{<\pt^{2}>}$ at desired invariant mass regions
in $70-90\%$ Pb--Pb collisions at $\snn=5.02$ TeV. The relevant data are presented as well for comparison.}
\label{tab:AvgPtSqu_Cmp}
\end{table*}

\begin{figure}[!htbp]
\begin{center}
\includegraphics[width=.40\textwidth]{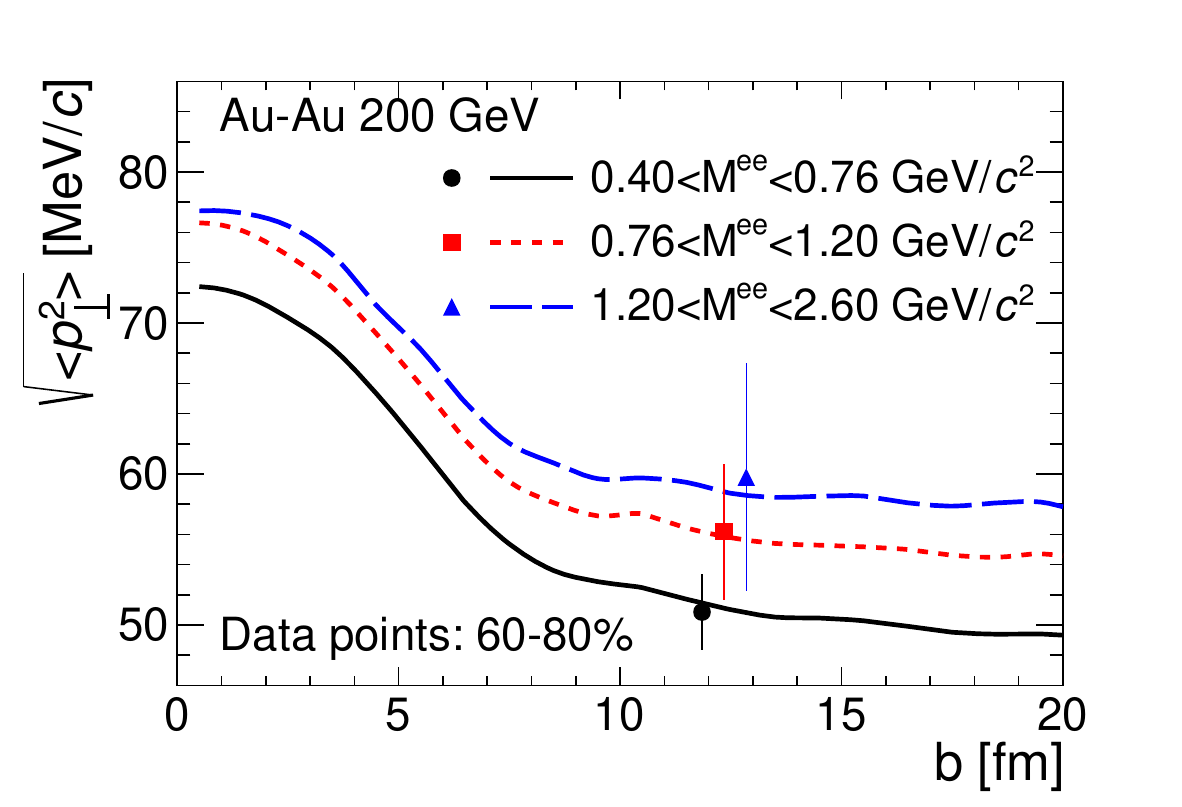}
\caption{(Color online) The dielectron pair-$\sqrt{<\pt^{2}>}$ distribution as a function of impact parameter
from various mass regions:
$0.4<M^{ee}<0.76~{\rm GeV}/{\it c}^{2}$ (solid black curve), $0.76<M^{ee}<1.2~{\rm GeV}/{\it c}^{2}$ (dotted red curve)
and $1.2<M^{ee}<2.6~{\rm GeV}/{\it c}^{2}$ (dashed blue curve).
The relevant measurements performed in $60-80\%$ Au--Au collisions at $\snn=200$ GeV~\cite{PhysRevLett.121.132301}
are shown as well for comparison.
For visibility, the data points are shifted horizontally.}
\label{fig:RHIC_AvgPtvsB}
\end{center}
\end{figure}

To investigate further the broadening of $\pt$ via $\sqrt{<\pt^{2}>}$,
we study its impact parameter $b$ dependence, as displayed in Fig.~\ref{fig:RHIC_AvgPtvsB}.
It can be seen that (1) $\sqrt{<\pt^{2}>}$ increases with decreasing $b$
and reaches a maximum value 1.5 times larger than that
in ultra-peripheral collisions with $b\approx 2R_{Au}\approx 13~fm$.
This result supports the statement~\cite{ALICE:2022hvk}
that the $\pt$ broadening originates
predominantly from the initial electromagnetic field strength
that varies significantly with impact parameters;
(2) moreover, as discussed in Fig.~\ref{fig:RHIC_AvgPtvsM},
the broadening depends on the invariant mass of the electron-positron pair,
behaving with an increasing trend with $M^{ee}$.
Similar results can be found in Refs.~\cite{Zha:2018tlq, PhysRevD.106.034025}.
In this figure, the STAR measurements~\cite{PhysRevLett.121.132301} are also plotted
for comparison and show good agreement within uncertainties.

\begin{figure*}[!htbp]
\begin{center}
\includegraphics[width=0.90\textwidth]{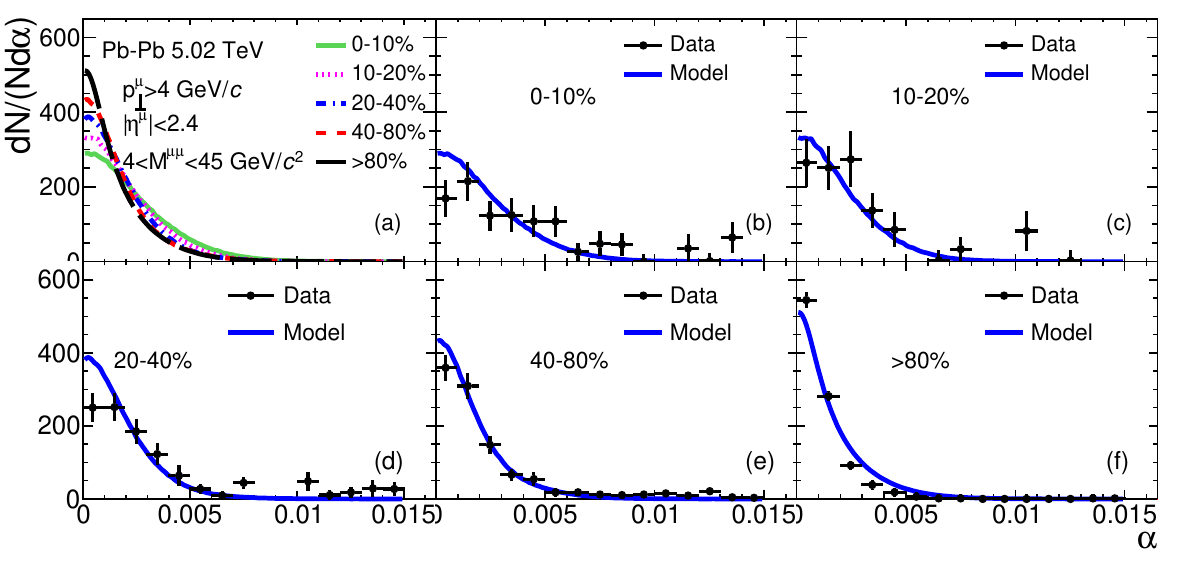}
\caption{(Color online) The dimuon pair-$\alpha$ distribution for the kinematics of the ATLAS measurements~\cite{PhysRevLett.121.212301}.
The panel-a (upper left) shows the model predictions within the different centrality classes with the curves in different styles,
while the other ones display the comparisons with the relevant experimental data (solid black point).}
\label{fig:ATLAS_PRL_dNdalpha}
\end{center}
\end{figure*}

With higher beam energy and intensity, the ATLAS Collaboration utilizes the dimuon pair acoplanarity, $\alpha$,
to characterize the $\pt$ broadening effects.
\begin{equation}
        \begin{aligned}\label{eq:acoplanery}
                \alpha&\equiv 1 - \frac{|\phi^{\mu^{+}}-\phi^{\mu^{-}}|}{\pi},
        \end{aligned}
\end{equation}
where, $\phi^{\mu^{\pm}}$ are the azimuthal angles of the single muons.
The model predictions are shown in Fig.~\ref{fig:ATLAS_PRL_dNdalpha}
with each distribution normalized to unity over the $\alpha$
range of interest, $\int d\alpha(\frac{1}{N}\frac{dN}{d\alpha})=1$.
The results are also filtered with the fiducial cuts
($\pt^{\mu}>4~{\rm GeV}/{\it c}$, $|\eta^{\mu}|<2.4$
and $4<M^{\mu\mu}<45~{\rm GeV}/{\it c}^{2}$) to allow a direct
comparison with the ATLAS measurements~\cite{PhysRevLett.121.212301}.
The panel-a in Fig.~\ref{fig:ATLAS_PRL_dNdalpha}
presents the model calculations within various centrality classes.
We observe that the $\alpha$ distributions are narrower in peripheral collisions
than those from more central collisions,
indicating that the transverse momentum broadening effect is more pronounced for central collisions,
where the impact parameters are small.
The same conclusion was found in Fig.~\ref{fig:RHIC_AvgPtvsB}.
This centrality-dependent broadening behavior is compared with
the available data for different centrality classes, as shown from panel-b to panel-f in Fig.~\ref{fig:ATLAS_PRL_dNdalpha}.
Within the experimental uncertainties,
our calculations can describe well the measured $\alpha$ dependencies.
We note that, even though the statistics of current ATLAS data is limited,
a centrality-dependent depletion of the dimuon yield is observed at small $\alpha$,
where the discrepancy between model and data is visible.
Such behavior may be due to missing effects such as the higher-order QED correction,
which will enhance (about 15$\%$ at maximum) the lepton pair transverse momentum broadening
in heavy-ion collisions with nuclear overlap~\cite{Li:2023yjt}.
Furthermore, the background contributions from dielectrons coming from
hadronic collisions (the ``hadronic cocktail'' subtracted in experimental measurements with nuclear overlap)
are more pronounced at low mass,
while the ones from the thermal medium radiation dominate in the intermediate mass region~\cite{RAPP2016586}.
These effects become important in particular in more central collisions~\cite{KLUSEKGAWENDA2019339, Zha:2018ywo}.

\begin{figure*}[!htbp]
\begin{center}
\includegraphics[width=0.40\textwidth]{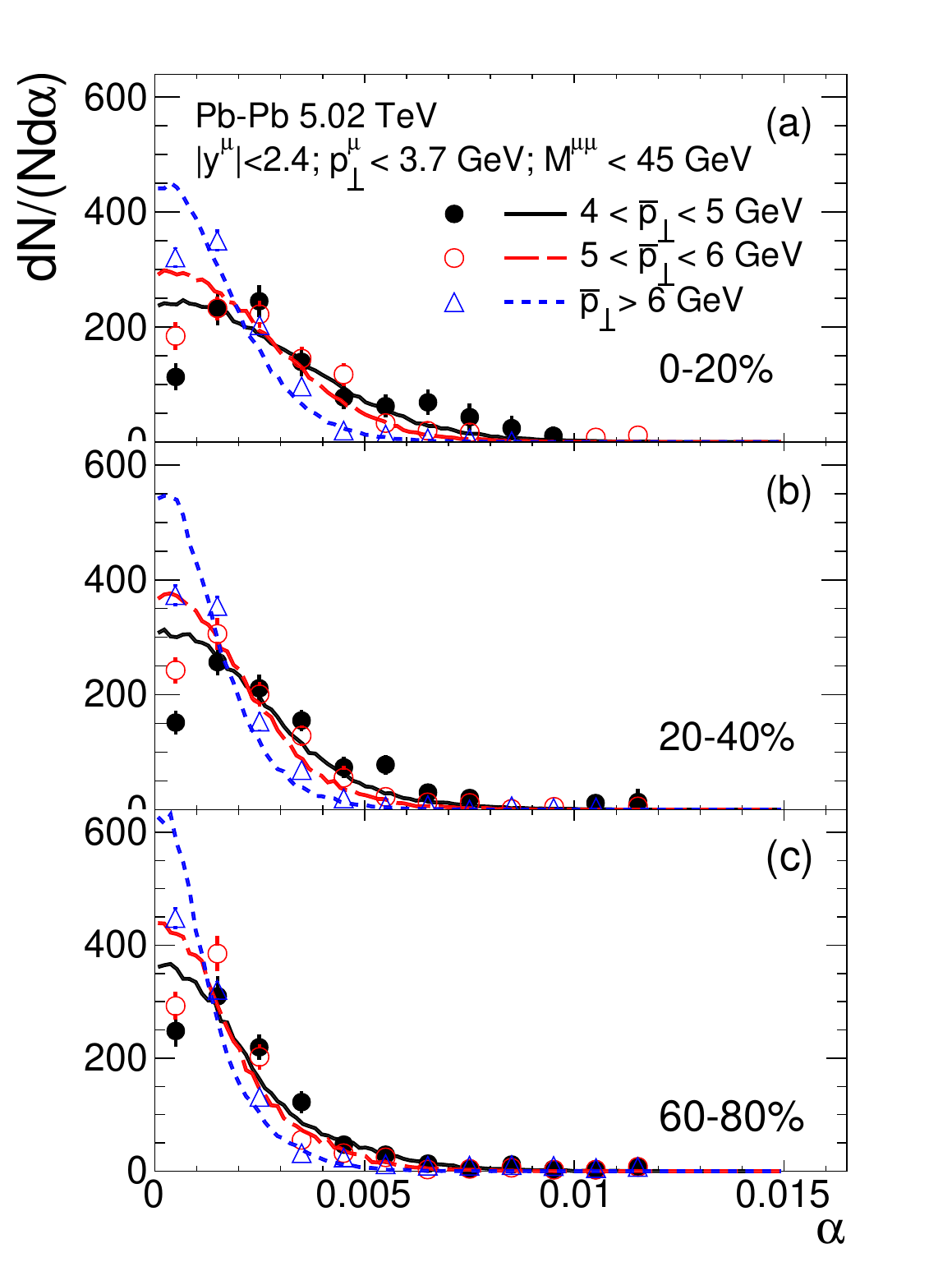}
\includegraphics[width=0.40\textwidth]{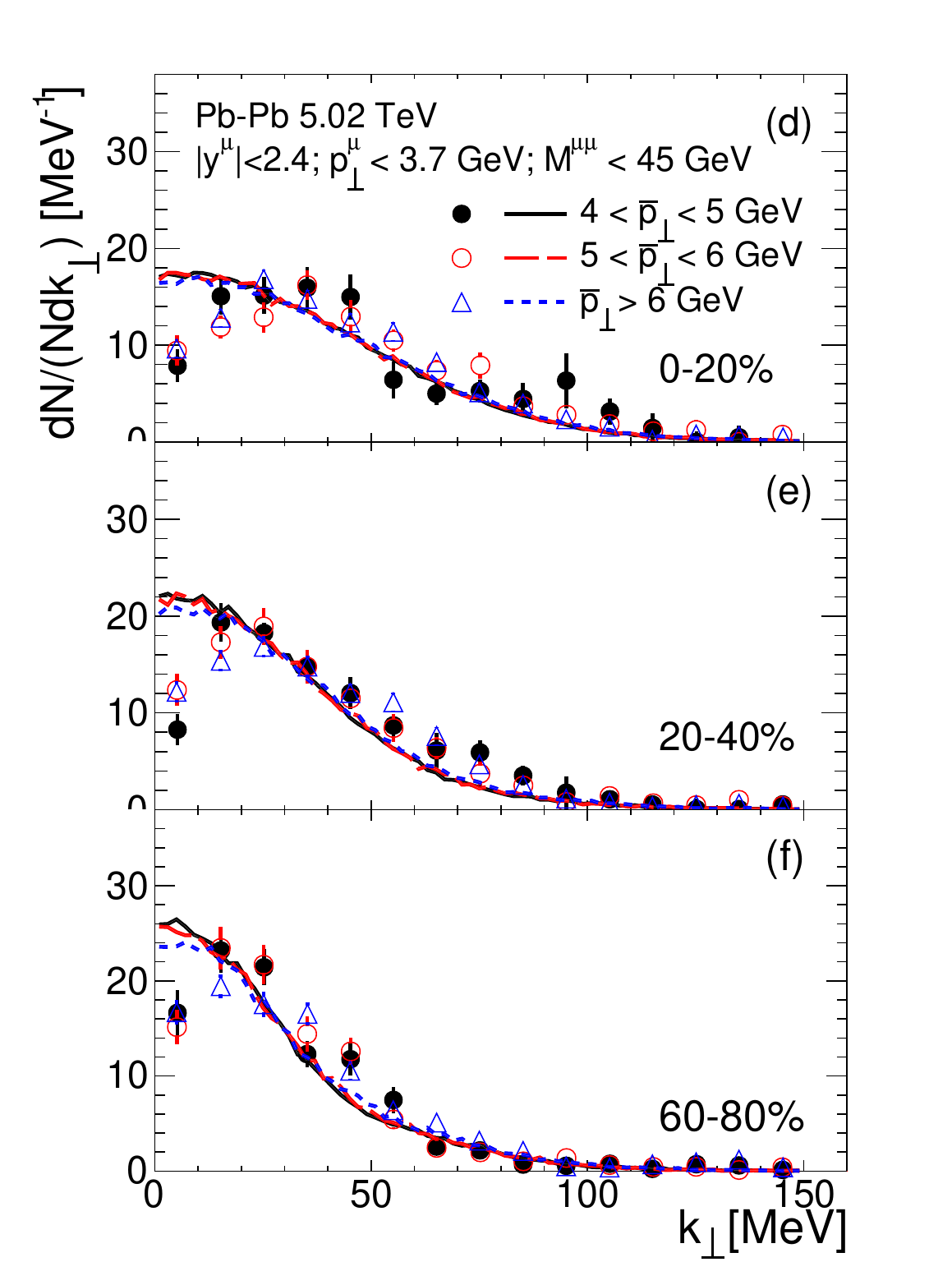}
\caption{(Color online) The dimuon pair-$\alpha$ (left-hand panels) and $k_{\perp}$ (right-hand panels) distributions within different
$\bar{p}_{\perp}$ intervals, $4<\bar{p}_{\perp}<5~\rm GeV$ (solid black curve),
$5<\bar{p}_{\perp}<6~\rm GeV$ (long dashed red curve), $\bar{p}_{\perp}>6~\rm GeV$ (dashed blue curve),
for three centrality intervals: $0-20\%$ (upper), $20-40\%$ (middle) and $60-80\%$ (bottom).
The relevant experimental data~\cite{PhysRevC.107.054907} (points) are presented as well for comparison.}
\label{fig:ATLAS_PRC_dNdalpha}
\end{center}
\end{figure*}
The centrality-dependent broadening of the $\alpha$ distributions is further studied according to the higher luminosity
measurement performed by ATLAS Collaboration~\cite{PhysRevC.107.054907}.
Meanwhile, it allows us to investigate the $\alpha$ broadening behavior within different average transverse momentum $\bar{p}_{\perp}$,
\begin{equation}
        \begin{aligned}\label{eq:AvgPt}
                \bar{p}_{\perp}&\equiv \frac{\pt^{\mu^{+}}+\pt^{\mu^{-}}}{2},
        \end{aligned}
\end{equation}
of the muons in the pair, where, $\pt^{\mu^{\pm}}$ are the transverse momentum of single muons.
We shown in the left panels of Fig.~\ref{fig:ATLAS_PRC_dNdalpha}
the dimuon pair-$\alpha$ distributions within three $\bar{p}_{\perp}$ ranges,
$4<\bar{p}_{\perp}<5~\rm GeV$ (solid black curve),  $5<\bar{p}_{\perp}<6~\rm GeV$ (long dashed red curve), and $\bar{p}_{\perp}>6~\rm GeV$ (dashed blue curve),
for three centrality intervals, $0-20\%$ (panel-a), $20-40\%$ (panel-b) and $60-80\%$ (panel-c).
The relevant data (points) are shown as well for comparisons.
We can see that: (1) the centrality-dependent broadening of the $\alpha$ distributions is confirmed,
and for a given $\bar{p}_{\perp}$ range, the broadening is more pronounced in central collisions, as observed in Fig.~\ref{fig:ATLAS_PRL_dNdalpha};
(2) the $\alpha$ broadening seems to have a visible $\bar{p}_{\perp}$-dependence with a considerable decrease from lower toward higher $\bar{p}_{\perp}$;
(3) the $\alpha$ distributions shown less depletion at $\alpha\approx0$ at higher $\bar{p}_{\perp}$,
which may be attributed to the quantum interference effects~\cite{Zha:2018tlq};
(4) for a given $\bar{p}_{\perp}$ range the depletion at small $\alpha$ becomes more significant in peripheral collisions
when comparing with that in central collisions, i.e. there is a centrality-dependent depletion at small $\alpha$;
furthermore, such depletion behavior is underestimated by the relevant calculations.
Consequently, the broadening of the $\alpha$ distributions are affected by both the centrality and the average transverse momentum.
It is argued~\cite{PhysRevC.107.054907} that the distributions of the transverse momentum scale $k_{\perp}$,
\begin{equation}
        \begin{aligned}\label{eq:kt}
                k_{\perp}&\equiv \pi\alpha\bar{p}_{\perp} = \frac{\pt^{\mu^{+}}+\pt^{\mu^{-}}}{2} \bigr( \pi-|\phi^{\mu^{+}}-\phi^{\mu^{-}}| \bigr),
        \end{aligned}
\end{equation}
which effectively quantify the component of the dimuon $\vec{p}_{\perp}$,
show no significant dependence on $\bar{p}_{\perp}$.
The relevant results are presented in the right panels of Fig.~\ref{fig:ATLAS_PRC_dNdalpha},
the dimuon pair-$k_{\perp}$ distributions within three $\bar{p}_{\perp}$ ranges,
$4<\bar{p}_{\perp}<5~\rm GeV$ (solid black curve),  $5<\bar{p}_{\perp}<6~\rm GeV$ (long dashed red curve), and $\bar{p}_{\perp}>6~\rm GeV$ (dashed blue curve),
for three centrality intervals, $0-20\%$ (panel-d), $20-40\%$ (panel-e) and $60-80\%$ (panel-f).
The relevant data (points) are shown as well for comparisons.
We can see that: (1) in a given centrality range, the $k_{\perp}$ distributions show a weak
$\bar{p}_{\perp}$-dependence from lower toward higher $\bar{p}_{\perp}$,
confirming that $k_{\perp}$ distributions have trivial $\bar{p}_{\perp}$-dependence;
(2) the $k_{\perp}$ broadening is more pronounced in central collisions;
(3) within the experimental uncertainties, the data can be well described by the relevant model calculations,
in particular at $k_{\perp}\gtrsim 20$ MeV.
Similarly to the $\alpha$ distributions (left panels),
the centrality-dependent depletion is also observed at small $k_{\perp}$ in the data,
which is underestimated by the model calculations.
Thus, the $k_{\perp}$ variable is better suited for assessing the centrality-dependent modifications of the dimuon alignment,
even though the physical interpretations of the depletion behavior at small $k_{\perp}$
are still challenging in general~\cite{PhysRevLett.121.212301, PhysRevC.107.054907}.
We plan to explore the underlying mechanisms in the future.

\section{Summary}\label{sec:summary}
In this work we have studied the photoproduction of lepton pair
in high energy nuclear collisions by employing a theoretical model with the equivalent photon approximation,
which can be derived from a full QED calculation at leading order in $\alpha_{\rm EM}$~\cite{Vidovic:1992ik}.
In the referenced work, the form factor is the one for a homogeneously charged sphere.
Moreover, the relevant calculations, such as the production cross section,
are obtained by integrating over the full phase space of the single leptons.

We update the model by taking into account the widely used Woods-Saxon distribution to
reproduce the nuclear profile more realistically.
Furthermore, to filter with the fiducial cut on the decaying leptons as applied in the current experimental measurements
and allow direct comparisons, we propose a Monte Carlo based strategy.
The calculated results can describe well the centrality-dependent broadening of both the acoplanarity ($\alpha$) and
the transverse momentum scale ($k_{\perp}$) distributions,
which are measured at the highest RHIC and LHC energies.
It is interesting to note that the $\alpha$ distributions vary with the average transverse momentum $\bar{p}_{\perp}$ of muons in the pair,
while the $k_{\perp}$ distributions do not.
Therefore, the $k_{\perp}$ variable is preferred for the study of the centrality-dependent broadening effect.
We also examine the updated model by calculating the
differential spectra as functions of pair mass $M^{ee}$ and the transverse momentum squared $\pt^{2}$,
and it is clearly found that the relevant experimental data can be well described,
indicating that the employed model is powerful for
characterizing the photoproduction of lepton pairs in ultra-relativistic heavy-ion collisions.


\section{Acknowledgements}\label{acknowledgements}
The authors are grateful to Prof. Jinfeng Liao and Prof. Wangmei Zha for helpful discussions and communications.
This work is supported by the National Natural Science Foundation of China (NSFC) under Grant No.12375137 and No.12005114.

\appendix

\section{Derivation of the energy flux of the photons in equivalent photon approximation (EPA)}\label{app:appendix}

\setcounter{equation}{0}
\renewcommand\theequation{A\arabic{equation}}

For the d'Alembert's equation (Eq.~\ref{eq:dAlbertEq}),
one can perform the Fourier transformation from momentum to position space, resulting in~\cite{KRAUSS1997503}
\begin{equation}\label{eq:A3}
A^{\nu}(k) = -\frac{1}{k^{2}}J^{\nu}(k),
\end{equation}
where, the current density reads
\begin{equation}
        \begin{aligned}\label{eq:A6}
                &J^{\nu}(k) = 2\pi Ze~ \delta(ku)~\rho(\sqrt{-k^{2}})~u^{\nu}~ e^{i\vec{k}\cdot \vec{b}},
        \end{aligned}
\end{equation}
with $u^{\nu}=\gamma(1,\vec{0}_{\perp},v)$ and $k=(\omega,\vec{k}_{\perp},k_{z})$.
$\rho$ is the nuclear charge density function.
Here we have chosen the $z$-axis as the direction of the Lorentz-boost $\gamma=1/\sqrt{1-v^2}$ between the two frames.
With the $\delta$-function in Eq.~\ref{eq:A6}, we have $-k^{2}=\vec{k}_{\perp}^{2} + (\frac{\omega}{\gamma})^{2}$.

The electromagnetic potential can be obtained by inserting Eq.~\ref{eq:A6} into Eq.~\ref{eq:A3}, yielding
\begin{equation}
        \begin{aligned}\label{eq:A8}
                A^{\nu}(k) &=2\pi Ze~ \delta(ku)~ \frac{\mathcal{F}(\sqrt{{-k^{2}}})}{-k^{2}}~u^{\nu}~ e^{i\vec{k}\cdot \vec{b}},
        \end{aligned}
\end{equation}
where, $Z$ is the nuclear charge number and $\mathcal{F}$
is the relevant normalized form factor defined as the Fourier transform of the charge distribution.

The electromagnetic field strength tensor in momentum space reads~\cite{KRAUSS1997503}
\begin{equation}
        \begin{aligned}\label{eq:B2}
                F^{\mu\nu}(k) &= -i{\biggr[}k^{\mu}A^{\nu}(k) - ik^{\nu}A^{\mu}(k) {\biggr]} \\
                &= \left( \begin{array}{cccc}
                        0     & -E_{x}   & -E_{y}   & -E_{z} \\
                        E_{x} &  ~~0     & -B_{z}   &  ~~B_{y} \\
                        E_{y} &  ~~B_{z} &  ~~0     & -B_{x} \\
                        E_{z} & -B_{y}   &  ~~B_{x} &  ~~0
                \end{array} \right).
        \end{aligned}
\end{equation}
Note that
\begin{itemize}
        \item[(1)] the transverse components of the electric field is given by
        \begin{equation}
                \begin{aligned}\label{eq:B3}
                        &\vec{E}_{\perp} = (F^{10},F^{20}) = -i {\bigr[} A^{0}\vec{k}_{\perp} - k^{0}\vec{A}_{\perp} {\bigr]} = -iA^{0}\vec{k}_{\perp}
            \end{aligned}
    \end{equation}
since $\vec{u}_{\perp}=\vec{A}_{\perp}=0$~(Eq.~\ref{eq:A8});
the transverse components of the magnetic field read
\begin{equation}
        \begin{aligned}\label{eq:B4}
                \vec{B}_{\perp} = \vec{v} \times E_{\perp} &= -ivA^{0}~(-k_{y},k_{x},0)^{T}.
        \end{aligned}
\end{equation}

\item[(2)] the longitudinal component of the electric field is
        \begin{equation}
                        \begin{aligned}\label{eq:B5}
                                &E_{z} = F^{30} = -i {\bigr[} k^{3}A^{0} - k^{0}A^{3} {\bigr]} \doteq 0,
                        \end{aligned}
        \end{equation}
since $k^{3}\doteq k^{0}=\omega$ and $A^{0}\doteq A^{3}$ in the ultra-relativistic limit ($v\approx c = 1$);
consequently, the longitudinal components of the electromagnetic fields vanish in this limit, resulting in the relations as
\begin{equation}
        \begin{aligned}\label{eq:B6}
                &\vec{E} \doteq \vec{E}_{\perp}, ~~\vec{B} \doteq \vec{B}_{\perp}, \\
                &|\vec{E}_{\perp}|\doteq|\vec{B}_{\perp}|,~~E_{z}\doteq B_{z}\doteq 0,~~|\vec{E}|\doteq |\vec{B}| \\
                &\vec{E} \perp \vec{B},~~\vec{E} \perp \vec{v},~~\vec{B} \perp \vec{v};
        \end{aligned}
\end{equation}

\item[(3)] the energy flux of the electromagnetic field through a plane perpendicular to the direction of the moving nucleus, is provided by the Poynting vector
\begin{equation}
        \begin{aligned}\label{eq:B7}
                &\vec{S}(\vec{r},t) \equiv \vec{E}(\vec{r},t) \times \vec{B}(\vec{r},t) \doteq {\bigr|} \vec{E}_{\perp}(\vec{r},t) {\bigr|}^{2}
        \end{aligned}
\end{equation}
in the limit $v\approx c=1$.
\end{itemize}

As discussed above, the electromagnetic field of a charged nucleus moving at high velocities becomes more and more
transverse with reference to the direction of propagation; see Eq.~\ref{eq:B6}.
As a consequence, an observer in the laboratory cannot distinguish between the electromagnetic fields of a relativistic nucleus
and a bunch of equivalent photons.
To extract the equivalent photon spectrum $n(\omega,\vec{x}_{\perp})$ which depends on the photon frequency $\omega$
and the displacement $\vec{x}_{\perp}$ in the transverse plane,
we can require that the energy flux of the electromagnetic fields through a transverse plane is
identical to the energy flux of the equivalent photons~\cite{KRAUSS1997503}, i.e., equivalent photon approximation (EPA):
\begin{equation}
        \begin{aligned}\label{eq:C1}
                &\int^{\infty}_{-\infty}dt \int^{\infty}_{-\infty}d^{2}\vec{x}_{\perp}\vec{S}(\vec{r},t) \equiv
                \int^{\infty}_{0}d\omega~\omega\int^{\infty}_{-\infty}d^{2}\vec{x}_{\perp}n(\omega,\vec{x}_{\perp}).
        \end{aligned}
\end{equation}
With Eq.~\ref{eq:B7}, the left hand side (LHS) of Eq.~\ref{eq:C1} can be expressed as
\begin{equation}
        \begin{aligned}\label{eq:C2}
                LHS &=\frac{1}{\pi}\int^{\infty}_{0}d\omega \int^{\infty}_{-\infty}d^{2}\vec{x}_{\perp} {\bigr|}\vec{E}_{\perp}(\vec{r},\omega){\bigr|}^{2}
        \end{aligned}
\end{equation}
Compared with the right hand side (RHS) of Eq.~\ref{eq:C1}, we have
\begin{equation}
        \begin{aligned}\label{eq:C3}
                &n(\omega,\vec{x}_{\perp}) = \frac{1}{\pi\omega} {\bigr|} \vec{E}_{\perp}(\vec{r},\omega) {\bigr|}^{2}
        \end{aligned}
\end{equation}
in the ultra-relativistic limit $v\approx c=1$.
The electric field in Eq.~\ref{eq:C3} is given by
\begin{equation}
        \begin{aligned}\label{eq:C4}
                {\bigr|}\vec{E}_{\perp}(\vec{r},\omega){\bigr|} &\stackrel{(\ref{eq:A8},\ref{eq:B3})}{=} \frac{Ze}{2\pi} {\biggr|} \int^{\infty}_{0}dk_{\perp}~k_{\perp}^{2}~\frac{\mathcal{F}(\sqrt{{-k^{2}}})}{-k^{2}}~J_{1}(x_{\perp}k_{\perp}) {\biggr|}
        \end{aligned}
\end{equation}
with the Bessel function
\begin{equation}
        \begin{aligned}\label{eq:C5}
                &J_{n}(z)\equiv \frac{1}{2\pi i^{n}}\int^{2\pi}_{0}d\theta e^{izcos\theta}e^{in\theta}
        \end{aligned}
\end{equation}
of the first kind ($n=1$).
The energy flux of the equivalent photons can be obtained by inserting Eq.~\ref{eq:C4} into Eq.~\ref{eq:C3}, yielding
\begin{equation}
        \begin{aligned}\label{eq:C6}
                &n(\omega,\vec{x}_{\perp}) = \frac{Z^{2}\alpha_{EM}}{\pi^{2}\omega} {\biggr|} \int^{\infty}_{0}dk_{\perp} k_{\perp}^{2} \frac{\mathcal{F}(\sqrt{{-k^{2}}})}{-k^{2}} J_{1}(x_{\perp}k_{\perp}) {\biggr|}^{2}
        \end{aligned}
\end{equation}
with the electromagnetic coupling factor $\alpha_{EM}=e^{2}/(4\pi)$.
Similar results can be found in Ref.~\cite{Vidovic:1992ik}.

Note that
\begin{itemize}
\item[(1)] the convergence of the integration part in Eq.~\ref{eq:C6},
is guaranteed if the form factor vanishes fast enough to zero at $k_{\perp}\rightarrow\infty$~\cite{Vidovic:1992ik};
we take $k^2_{\perp} \leqslant \frac{1}{R^{2}} - {\bigr(}\frac{\omega}{\gamma}{\bigr)}^{2}$ in the numerical calculations~\cite{Niu:2022cug},
where $R$ is the nuclear radius of the heavy-ion, see Eq.~\ref{eq:WoodsSaxon};

\item[(2)] the transverse plane integrated energy flux in Eq.~\ref{eq:C3}, can be obtained via
\begin{equation}
        \begin{aligned}\label{eq:C8}
                n(\omega) &= \int^{\infty}_{-\infty}d^{2}\vec{x}_{\perp}n(\omega,\vec{x}_{\perp}) \\
                &\stackrel{(\ref{eq:B3},\ref{eq:A8})}{=}\frac{2Z^{2}\alpha_{EM}}{\pi\omega} \int^{\infty}_{0}dk_{\perp} |\vec{k}_{\perp}|^{3} {\biggr[} \frac{\mathcal{F}(\sqrt{{-k^{2}}})}{-k^{2}} {\biggr]}^{2},
        \end{aligned}
\end{equation}
resulting in
\begin{equation}
        \begin{aligned}\label{eq:C9}
                &\frac{dn(\omega,k_{\perp})}{dk_{\perp}} = \frac{2Z^{2}\alpha_{EM}}{\pi\omega} |\vec{k}_{\perp}|^{3} {\biggr[} \frac{\mathcal{F}(\sqrt{{-k^{2}}})}{-k^{2}} {\biggr]}^{2}.
        \end{aligned}
\end{equation}

\end{itemize}

\bibliography{EM}
\end{document}